\documentclass[a4paper]{article}
\usepackage{graphicx}

 \newcommand{\lan}{\langle}
 \newcommand{\ran}{\rangle}
 \newcommand{\be}{\begin{equation}}
  \newcommand{\bea}{\begin{eqnarray}}
\newcommand{\eea}{\end{eqnarray}}
  \newcommand{\ee}{\end{equation}}

\def\la{\mathrel{\mathpalette\fun <}}

\def\fun#1#2{\lower3.6pt\vbox{\baselineskip0pt\lineskip.9pt
\ialign{$\mathsurround=0pt#1\hfil##\hfil$\crcr#2\crcr\sim\crcr}}}
\newcommand{\vep}{\mbox{\boldmath${\rm p}$}}

\title{Nonperturbative vacuum and condensates in QCD
below thermal phase transition \\ }
\author{N.O.Agasian\thanks{ E-mail: {\tt agasian@heron.itep.ru}}
\\
\\ Institute of Theoretical and Experimental Physics,
\\Moscow 117218, Russia\\}

\date{}

\begin{document}

\maketitle
\begin{abstract}

Thermodynamic properties  of the QCD nonperturbative vacuum with
two light quarks are studied. It is shown that at low temperatures
$T\la M_\pi$  relativistic massive pions can be treated within the
dilute gas approximation. Analytic temperature dependence of the
quark condensate is found in perfect agreement with the numerical
calculations obtained at the three-loop level of the chiral
perturbation theory with non-zero quark mass. The gluon condensate
slightly varies with the increase of the temperature. It is shown
that the temperature derivatives of the anomalous and normal
(quark massive term) contributions to the trace of the
energy-momentum tensor in  QCD  are equal to each other in the low
temperature region.

 \end{abstract}
 \vspace{2cm}


\newpage

 1. The investigation of the vacuum state behavior under the
 influence of various external factors is known to be one of the
 central problems of quantum field theory. In the realm of strong
 interactions (QCD) the main factors are the temperature and
 the baryon density. At low temperatures, $T<T_c$ ( $T_c$--temperature of the
 "hadron--quark-gluon" phase transition ), the dynamics of QCD
 is essentially nonperturbative and
 is characterized by confinement and spontaneous
breaking of chiral symmetry (SBCS). In the hadronic phase the
partition function of the system is dominated by the contribution
 of the lightest  particles in the physical spectrum. In QCD this
role is played by the $\pi $-meson which is the Goldstone
excitation mode in chiral condensate in the limit of two massless
flavors. Therefore the low temperature physics can be described in
terms of the effective chiral theory \cite{1,2,3}. An important
problem is the behavior of the order parameter (the quark
condensate $\langle \bar q q\rangle)$ with the increase of the
temperature.  In the ideal gas approximation the contribution of
the massless pions into the quark condensate is proportional to $
T^2$ \cite{4,5}. In chiral perturbation theory (ChPT) the two --
and three--loop contributions ($\sim T^4$  and  $\sim T^6$
 correspondingly) into $\langle \bar q q \rangle$ have been found in
\cite{5}  and \cite{6,7}.In the general case of $N_f$ massless
flavors, the low temperature expansion of the quark condensate
takes the form
 \be
\frac{\lan\bar qq\ran(T)}{\lan \bar q q
\ran}=1-\frac{(N^2_f-1)}{N_f} \frac{T^2}{12 F^2_\pi}
-\frac{(N^2_f-1)}{2N^2_f} \left( \frac{T^2}{12
F^2_\pi}\right)^2-N_f(N^2_f-1)
 \left( \frac{T^2}{12
F^2_\pi}\right)^3\ln \frac{\Lambda_q}{T}+O(T^8),
 \label{1}
 \ee
where $F_\pi\simeq93 $MeV is the value of the pion decay constant
and $\Lambda_q\simeq470$Mev ($N_f=2$)\cite{7}. The result
(\ref{1}) is an exact theorem of QCD in chiral limit.

 The situation  with the gluon condensate $\langle
G^2\rangle\equiv \langle (gG^a_{\mu\nu})^2\rangle$is very
different. The gluon condensate is not an order parameter of the
phase transition and it does not lead to any spontaneous symmetry
breaking (SSB). At the quantum level the trace anomaly leads to
the breaking of the scale invariance  which is phenomenologically
described by non-zero value of $\langle G^2 \rangle$.
However, this is not a SSB phenomenon and hence does not lead to
the appearance of the Goldstone particle. The mass of the lowest
excitation (dilaton)is directly connected to the gluon condensate,
$ m_D\propto (\langle G^2\rangle)^{1/4}$.
Thus, in the gluonic sector of QCD, the thermal excitations of
glueballs are exponentially suppressed by the Boltzmann factor
$\sim \exp \{-m_{gl}/T\}$ and their contribution to the shift of
the gluon condensate is small ($\Delta\langle G^2\rangle/\langle
G^2\rangle\sim $0.1 \% at $T\sim$200 MeV) \cite{8}.  Next we note
that in the one-loop approximation of ChPT pions are described as a
gas of massless noninteracting particles. Such a system is
obviously scale-invariant and therefore does not contribute to
the trace of the  energy-momentum tensor and correspondingly to
$\langle G^2\rangle$.  As it has been demonstrated in \cite{9}
the gluon condensate temperature dependence arises only at the
ChPT three--loop level due to the interaction among the Goldstone
bosons.

For the quark condensate the account of the non-zero quark mass was
done in \cite{7}. Within ChPT two masses of the $u$ and $d$ quarks
are considered as perturbation ($m_u$ and $m_d$ are much smaller
than any scale of the theory). Correspondingly thermodynamic
quantities ( pressure, $\lan \bar q q\ran(T))$ are represented as
a series in powers of $T$ and $m$ in low temperature expansion. As a
consequence of the fact that the spectrum of the unperturbed
system contains massless pions, the quark mass term generates
infrared singularities involving negative powers of $T$ \cite{7}.
In Ref. \cite{7} results of a numerical procedure was presented
in the form of the graphical dependence of $\lan \bar q q\ran(T)$
as a function of $T$, which we will compare with our analytic
formula. The deviation of the quark condensate from
the chiral limit (\ref{1}) turns out to be  substantial
($\sim 25$\%) at $T=140$ MeV in both cases.

It is well known that due to the smallness of pion mass as
compared to the typical scale of strong interactions, the pion
plays a special role among other strongly-interacting particles.
Therefore for many problems of QCD at zero temperature the chiral
limit, $M_\pi\to 0$, is an appropriate one. On the other hand a
new mass scale emerges in the physics of QCD phase transitions,
namely the critical transition temperature $T_c$. Numerically the
critical transition temperature turns out to be close to the pion
mass, $T_c\approx M_\pi$.\footnote{The deconfining phase
transition temperature is the  one obtained in lattice
calculations $T_c(N_f=2)\simeq 173 $ MeV and $T_c(N_f=3)\simeq
154$ MeV \cite{10}.} Thus in the confining phase of QCD the small
parameter $M_\pi/T$ is lacking and hence the interval $M_\pi\ll
T<T_c$ where the high temperature expansion is applicable for
thermal pions does not exist. However hadron states heavier than
pion have masses several times larger than $T_c$ and therefore
their contribution to the thermodynamic quantities is damped by
Bolzmann factor $\sim \exp \{ -M_{hadr}/T\}$. Thus the
thermodynamics of the low temperature hadron phase, $T<M_\pi$, is
described basically in terms of the thermal  excitations of
relativistic massive pions.

In  the present paper we study the behavior of the quark and gluon
condensates in the low temperature phase. It is shown that the relativistic
dilute massive pion gas approximation is applicable at temperature
$T\la M_\pi$. Analytical expressions for condensates are obtained. The
 comparison with the results of three-loop ChPT calculations
 $\langle \bar q q \rangle(T)$ and $\langle G^2\rangle(T)$ is made.
 The low temperature relation
for the trace of the energy-momentum tensor in QCD with two light
quarks is obtained based on the general dimensional and
renormalization-group properties of the QCD partition function and
dominating role of the pion thermal exitations in the hadronic
phase.

 2. For non-zero quark mass ($m_q\neq 0)$ the scale invariance is
 broken already at the classical level. Therefore the pion thermal
 excitations would change, even in the ideal gas approximation, the
 value of the gluon condensate with increasing temperature.
 To determine this dependence can be use the general
 renormalization and scale properties of the QCD partition
 function. This is a standard method  and it is used for derivation of
 low-energy QCD theorems \cite{11}.
  For QCD at finite temperature and chemical potential
  these theorems were derived in \cite{12,13}.
 This method was used for investigation of QCD vacuum phase
 structure in a magnetic field \cite{14} and at finite temperature
 \cite{15} and also in nuclear matter \cite{Lee}.

 A relation between the trace anomaly and thermodynamic pressure in pure-glue
 QCD was obtained in \cite{Land} by making use of the dimensional regularization in
 the framework of the renormalization group (RG) method. Also within the RG method,
 but by employing slightly different techniques an analogous relation was
 derived in the theory with quarks in ref. \cite{ag1}.
 Then, by making use of the results obtained in the above-mentioned papers,
 we can write the following expression for the trace anomaly in QCD with quarks:

 \be
\lan \theta^g_{\mu\mu}\ran
=\frac{\beta(\alpha_s)}{16\pi\alpha_s^2} \langle G^2\rangle
=-(4-T\frac{\partial}{\partial
T}-\sum_q(1+\gamma_{m_q})m_q\frac{\partial}{\partial {m_q}}) P_R,
\label{10a}
 \ee
 where $P_R$ is the renormalized pressure, $\beta(\alpha_s)=d\alpha_s(M)/d
  ~ln M$  is the Gell-Mann-Low  function and $\gamma_{m_q}$ is the
anomalous dimension of the quark mass.
   It is convenient to choose such a large scale that one can take the
lowest order expressions,  $\beta(\alpha_s)\to -
b\alpha^2_s/2\pi$, where $b=(11 N_c-2N_f)/3$ and $1+\gamma_m\to
1$. Thus, we have the following equations for condensates
\be
\lan G^2\ran (T)=\frac{32\pi^2}{b} (4-T\frac{\partial}{\partial
T}-\sum_q m_q\frac{\partial}{\partial m_q}) P_R\equiv
 \hat DP_R~,
 \label{11}
 \ee
 \be
 \lan\bar q q\ran (T)=-\frac{\partial P_R}{\partial {m_q}}~.
 \label{12}
 \ee

 3. The density of the pionic gas at temperature $T$ is given by
 \be
 n_\pi(T)=3\int\frac{d^3p}{(2\pi)^3} \frac{1}{\exp(\sqrt{\vep^2+
 M^2_\pi}/T)-1}~,
 \label{13}
 \ee
 and the average distance between particles in this gas is
 $L_\pi\simeq n^{-1/3}_\pi$. The dilute gas approximation is
 valid, when the average distance $L_\pi$ is much smaller than
 the mean free path $\lambda_{\pi\pi}$ calculated taking into
 account collisions, $L_\pi\ll \lambda_{\pi\pi}$. For a hadronic
 gas, the mean free path is well known \cite{17, 18} at low
 temperatures, where the gas almost exclusively consists of pions.
 An analysis of the $\pi\pi$ collision rate shows that in the
 range 50 MeV$<T<$140 MeV, ChPT formula for mean free path
 $\lambda_{\pi\pi} \simeq 12 F^4_\pi/T^5$ \cite{18} is valid to
 within about 20\% \cite{19}. In Fig.1 the behavior of
 the ratio $L_\pi/\lambda_{\pi\pi}$ as a function of $T$ is shown.
Hence at $T\la M_\pi$ the gas approximation for pions can be used.

\begin{figure}[!ht]
\begin{picture}(320,190)
\put(0,15){\includegraphics{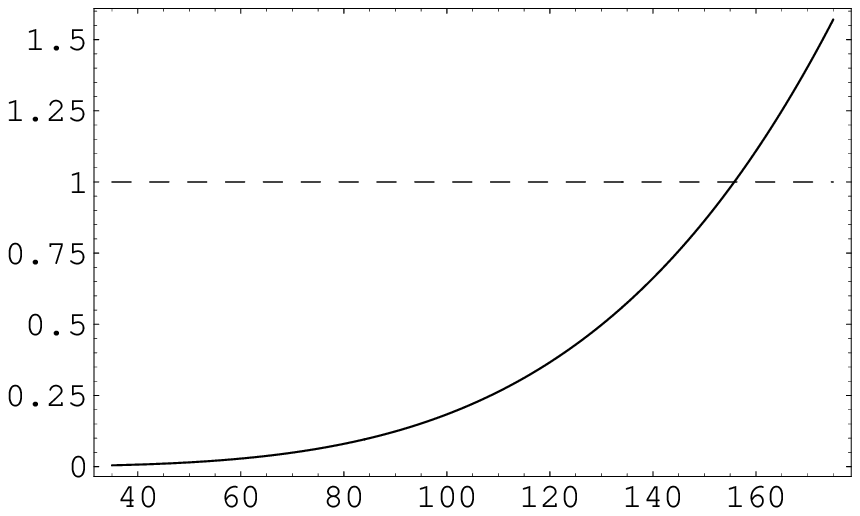}}
\put(0,90){\rotatebox{90}{$L_\pi/\lambda_{\pi\pi}$}}
\put(130,15){T (MeV)}
\put(100,120){}
\end{picture}
\caption{The ratio of the mean interparticle distance $L_\pi$ ( in pionic gas )
 to the mean free path  $\lambda_{\pi\pi}$ as a function of the temperature}
\label{Fig.1}
\end{figure}

Thus the effective pressure from which one can extract the
condensates $\lan \bar q q\ran(T)$ and $\lan G^2\ran(T)$ using
the relations (\ref{11}) and (\ref{12}) has the form
\be
P_{eff}(T)=-\varepsilon_{vac}+P_\pi(T),
\label{14}
\ee
where
$\varepsilon_{vac}=\frac14\lan\theta_{\mu\mu}\ran$ is the
nonperturbative vacuum energy density at $T=0$ and
\be
\lan
\theta_{\mu\mu}\ran=-\frac{b}{32\pi^2} \lan G^2\ran+\sum_{q=u,d}
m_q\lan\bar qq \ran
\label{15}
\ee
is the trace of the
energy-momentum tensor. In Eq.(\ref{14}) $P_\pi$ is the massive
pions pressure
\be
P_\pi=-3T\int\frac{d^3p}{(2\pi)^3}\ln
(1-e^{-\sqrt{\vep^2+M^2_\pi}/T}).
\label{16}
\ee
The quark and
gluon condensates are given by the equations
\be
\lan \bar qq\ran
(T)=-\frac{\partial P_{eff}}{\partial m_q},
\label{17}
\ee
\be
\lan G^2\ran (T)= \hat DP_{eff},
\label{18}
\ee
where the
operator $\hat D$ is defined by the relation (\ref{11})
\be \hat
D=\frac{32\pi^2}{b} (4-T\frac{\partial}{\partial T}-\sum_q
m_q\frac{\partial}{\partial m_q}).
\label{19}
\ee

Consider the $T=0$ case. One can use the low energy theorem
for the derivative of the gluon condensate with respect to the
quark mass \cite{11}
\be
\frac{\partial}{\partial m_q}\lan G^2\ran= \int d^4 x\lan G^2(0)
\bar q q(x)\ran =-\frac{96\pi^2}{b}\lan \bar q q\ran+O(m_q),
\label{20}
\ee
where $O(m_q)$ stands for the terms linear in light
quark masses.Then one arrives at the following
relation\footnote{The equation (\ref{21}) contains corrections
$\sim  m_q\partial\lan \bar qq\ran/\partial m_q \sim O(m_q)$
which are negligible for $u$ and $d$ quarks.}
\be
\frac{\partial\varepsilon_{vac}}{\partial m_q}=-
\frac{b}{128\pi^2}\frac{\partial}{\partial m_q} \lan
G^2\ran+\frac{1}{4}\lan \bar q q\ran =\frac34 \lan \bar q
q\ran+\frac14\lan \bar q q\ran=\lan \bar q q\ran.
\label{21}
\ee
Note that three fourths of the quark condensate stem from the
gluon part of the nonperturbative vacuum energy density. Along the
same lines one arrives at the following expression for the gluon
condensate
\be
-\hat
D\varepsilon_{vac}=-\frac{32\pi^2}{b}(4-\sum_q
m_q\frac{\partial}{\partial m_q})(-\frac{b_0}{128\pi^2}\lan
G^2\ran+\frac14 \sum_qm_q\lan \bar q q\ran) =\lan G^2\ran.
\label{22}
\ee

In order to get the dependence of the quark and
gluon condensates upon $T$ use is made of the Gell-Mann-
Oakes-Renner (GMOR) relation ($\Sigma=|\lan\bar u u\ran|=|\lan
\bar dd\ran|$)
\be F^2_\pi M^2_\pi=-\frac12(m_u+m_d)\lan \bar
uu+\bar dd\ran=(m_u+m_d)\Sigma
\label{23}
\ee
Then we can find the following relations
\be
\frac{\partial}{\partial
m_q}=\frac{\Sigma}{F^2_\pi} \frac{\partial}{\partial M^2_\pi},
\label{24}
\ee
\be \sum_qm_q\frac{\partial}{\partial
m_q}=(m_u+m_d)\frac{\Sigma}{F^2_\pi}\frac{\partial}{\partial
M^2_\pi}=M^2_\pi\frac{\partial}{\partial M^2_\pi},
\label{25}
\ee
\be \hat D=\frac{32\pi^2}{b}(4-T\frac{\partial}{\partial
T}-M^2_\pi\frac{\partial}{\partial M^2_\pi}).
\label{26}
\ee

The pressure of the massive relativistic pion gas has the form
$$
P_\pi(T)=-\frac{3T^4}{2\pi^2}\int^\infty_0 x^2
dx\ln(1-e^{-\omega_\pi(x)})=\frac{3T^4}{2\pi^2}
\sum^\infty_{n=1}\frac{1}{n}\int^\infty_0 x^2 dx
e^{-n\omega_\pi(x)}
$$
\be =\frac{3M^2_\pi T^2}{2\pi^2}
\sum^\infty_{n=1}\frac{1}{n^2}K_2(n\frac{M_\pi}{T}),
\label{27}
\ee
where $\omega_\pi (x)=\sqrt{x^2+M^2_\pi/T^2}$ and $K_2$ is the
Mackdonald function.
Making use of (\ref{17},\ref{21},\ref{24}) and (\ref{27})
one gets for the quark condensate
\be
\frac{\Sigma(T)}{\Sigma}=1-\frac{3T^2}{4\pi^2F^2_\pi}\int^\infty_0\frac{x^2
dx}{\omega_\pi (x)(e^{\omega_\pi (x)}-1)}=1-\frac{3M_\pi
T}{4\pi^2F^2_\pi}\sum^\infty_{n=1}
\frac{1}{n}K_1(n\frac{M_\pi}{T}).
\label{28}
\ee

\begin{figure}[!ht]
\begin{picture}(320,178)
\put(0,15){\includegraphics{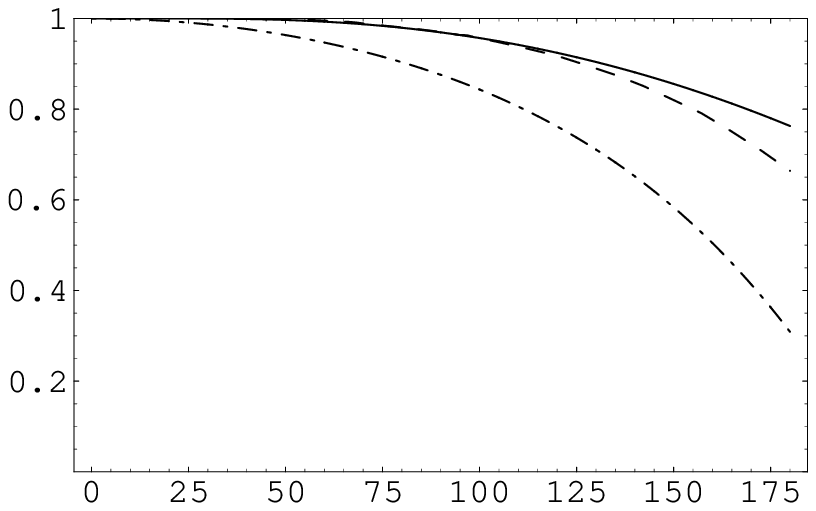}}
\put(0,90){\rotatebox{90}{$\Sigma(T)/\Sigma$}}
\put(130,15){T (MeV)}
\put(100,120){}
\end{picture}
\caption{The quark condensate  $\Sigma(T)/\Sigma$ as a function
of the temperature. The solid line corresponds to
Eq.(\ref{28}). The dash-dotted line corresponds to the
three-loop ChPT formula (\ref{1}) in chiral limit for $N_f=2$.
The dashed line is the numerical three-loop result of ChPT with
non-zero quark mass ( $M_\pi$=140 MeV ) from Ref.\cite{7}.}
\label{Fig.2}
\end{figure}

In the chiral limit, $M_\pi\ll T<T_c$, the shift of the quark condensate is
given by the standard temperature one-loop ChPT contribution to
$\lan \bar q q\ran(T)$

\be
\frac{\Delta \Sigma}{\Sigma}=-\frac{3T^2}{4\pi^2F^2_\pi}
\sum^\infty_{n=1} \frac{1}{n^2}=-\frac{T^2}{8F^2_\pi}.
\label{29}
\ee
In the opposite limit of low temperatures, $T\ll M_\pi$, one
obtains the following result
\be
\frac{\Delta \Sigma}{\Sigma}= -\frac{3M_\pi T}{4\pi^2F^2_\pi}
K_1(\frac{M_\pi}{T})\to -\frac{3M^{1/2}_\pi
T^{3/2}}{2^{5/2}\pi^{3/2} F^2_\pi}e^{-M_\pi/T}.
\label{30}
\ee

The value of the quark condensate as a function of the temperature
is depicted in Fig. 2. One can see that the gas formula (\ref{28})
for the relativistic massive pions perfectly agrees with the
numerical calculation of $\Sigma (T)$ \cite{7} within the framework
of ChPT with non-zero quark mass. At the temperature $T=M_\pi$ the
deviation is only 2.5\%.

4. Now let us consider the temperature dependence of the gluon
condensate within the framework of the present approach. Eqs.
(\ref{18}), (\ref{22}) and (\ref{26}) yield
\be
\frac{\lan G^2\ran
(T)}{\lan G^2\ran }=1-\frac{24}{b}\frac{M^2_\pi T^2}{\lan G^2\ran
}\int^\infty_0\frac{x^2dx}{\omega_\pi (x)(e^{\omega_\pi (x)}-1)}=1
-\frac{24}{b}\frac{M^3_\pi T}{\lan G^2\ran
}\sum^\infty_{n=1}\frac{1}{n}K_1(n\frac{M_\pi}{T}).
\label{31}
\ee
For $M_\pi\ll T$ one gets
\be
\frac{\Delta \lan G^2\ran }{\lan
G^2\ran }= -\frac{24}{b}\frac{M^2_\pi T^2}{\lan G^2\ran}
\sum^\infty_{n=1}\frac{1}{n^2}=-\frac{4\pi^2}{b} \frac{M^2_\pi
T^2}{\lan G^2\ran}.
\label{32}
\ee
In chiral limit, $M_\pi=0$, we have $\Delta \lan G^2\ran(T)=0$,
in agreement with the fact that a free gas of massless particles
is conformally invariant. In the opposite  nonrelativistic
(low temperature) limit, $T\ll M_\pi$, one obtains
\be
\frac{\Delta \lan G^2\ran }{\lan G^2\ran }=
-\frac{24}{b}\frac{M^3_\pi T}{\lan G^2\ran} K_1(\frac{M_\pi}{
T})\to- \frac{3\cdot2^{5/2}\pi^{1/2}}{b}\frac{M^{5/2}_\pi T^{3/2}}{\lan
G^2\ran} e^{-M_\pi/T}.
\label{33}
\ee
On the other hand, as stated
earlier, within ChPT and for zero quark mass the dependence of the
gluon condensate on the temperature arises only  at the three-loop
level due to the interaction of the massless pions \cite{9}. In
Refs. \cite{6,7} the pressure of the $N_f$ massless quarks has
been calculated at the three-loop level with the result
\be
P^{N_f=2}_{ChPT}=\frac{\pi^2}{30}T^4\{1+4(\frac{T^2}{12F^2_\pi})^2\ln
\frac{\Lambda_p}{T}\}+O(T^{10}),
\label{34}
\ee
where the
numerical value, relevant for the limit $m_u=m_d=0$ at fixed
$m_s$ is $\Lambda_p\simeq 275$ MeV \cite{7}. Then using
(\ref{18}), (\ref{19}) and (\ref{24}) one gets for the gluon
condensate the result of Leutwyler \cite{9}\footnote{In case
$M_\pi=0$ one has $(T \partial/\partial T-4)P=\varepsilon -3
P=\lan \theta^g_{\mu\mu}\ran_T$ \cite{9}.}
\be
\frac{ \lan G^2\ran
(T)_{ChPT} }{\lan G^2\ran }=1
-\frac{16\pi^4}{135b}\frac{T^8}{F^4_\pi\lan
G^2\ran}(\ln\frac{\Lambda_p}{T}-\frac{1}{4}).
 \label{35}
 \ee
In Fig.3 we show the temperature dependence of the gluon
condensate  at $\lan G^2\ran=0.5$ GeV$^4$.\footnote{This value
corresponds to the standard magnitude $\lan
\frac{\alpha_s}{\pi}F^2_{\mu\nu}\ran = 0.012$ GeV$^4$ \cite{20}.}

\begin{figure}[!ht]
\begin{picture}(320,190)
\put(0,15){\includegraphics{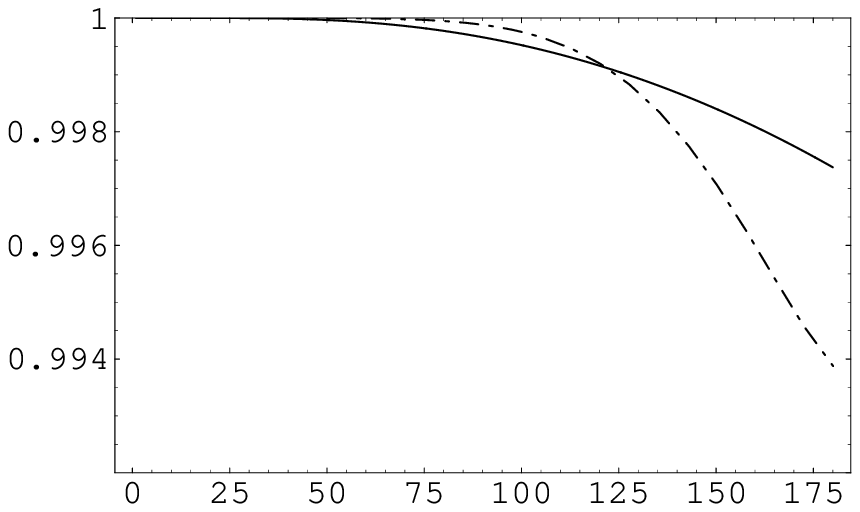}}
\put(0,75){\rotatebox{90}{$\lan G^2\ran (T)/\lan G^2\ran$}}
\put(130,15){T (MeV)}
\put(100,120){}
\end{picture}
\caption{The gluon condensate
$\lan G^2\ran (T)/\lan G^2\ran$ as a function of the temperature.
The solid line corresponds to Eq.(\ref{31}) at $\lan G^2\ran=0.5$
GeV$^4$. The dash-dotted line corresponds to the three-loop
result of ChPT in the chiral limit (Eq.(\ref{35})).}
\label{Fig.3}
\end{figure}

One can get an insight into the physical nature of small $\lan
G^2\ran$ shift with the increase of the temperature. The smallness
of this quantity is due to the large value of $\lan G^2\ran$ at
$T=0$ as compared to the typical parameters of the thermal
hadronic phase, $\Delta\lan G^2\ran/\lan G^2\ran \propto M^2_\pi
T^2_c/\lan G^2\ran \simeq 10^{-3}$ at $M_\pi=0.14$ GeV,
$T_c=0.17$ GeV and $\lan G^2\ran =0.5$ GeV$^4$.

5. Within the above framework one can derive the thermodynamic
relation for the quantum anomaly in the trace
 of the energy-momentum tensor of QCD. At low temperature the main
 contribution to the pressure comes from thermal excitations of
 massive pions. The general expression for the pressure reads

 \be
 P_\pi=T^4\varphi(M_\pi/T),
 \label{36}
 \ee
 where $\varphi$ is a function of the ratio $M_\pi/T$ and in case of the dilute
 gas it is given by equation (\ref{27}). Then the following
 relation is valid
 \be
 (4-T\frac{\partial}{\partial T}-M^2_\pi\frac{\partial}{\partial
 M^2_\pi}) P_\pi=M^2_\pi\frac{\partial P_\pi}{\partial M^2_\pi}.
 \label{37}
 \ee
Making use of (\ref{17},\ref{18}), (\ref{21},{22}) and (\ref{37})
one gets
\be
\Delta \lan \bar qq\ran =-\frac{\partial
P_\pi}{\partial m_q},~~ \Delta \lan G^2\ran=\frac{32\pi^2}{b}
M^2_\pi\frac{\partial P_\pi}{\partial M^2_\pi},
\label{38}
\ee
where $ \Delta \lan \bar qq\ran= \lan \bar qq\ran_T- \lan \bar
qq\ran $ and $\Delta \lan G^2\ran=  \lan G^2\ran_T- \lan G^2\ran.$
In view of (\ref{25}) one can recast (\ref{38}) in the form
\be
\Delta \lan G^2\ran=-\frac{32\pi^2}{b} \sum_q m_q\Delta  \lan \bar
qq\ran
\label{39}
\ee Let us divide both sides of (\ref{39}) by
$\Delta T$ and take the limit $\Delta T\to 0$. This yields
\be
\frac{\partial \lan G^2\ran}{\partial T}=-\frac{32\pi^2}{b} \sum_q
m_q\frac{\partial\lan \bar qq\ran}{\partial T}.
\label{40}
\ee
This can be rewritten as
\be
\frac{\partial \lan
\theta^g_{\mu\mu}\ran}{\partial T}=\frac{\partial \lan
\theta^q_{\mu\mu}\ran}{\partial T} \label{41}
\ee where $ \lan
\theta^q_{\mu\mu}\ran=\sum m_q \lan \bar qq\ran$ and $  \lan
\theta^g_{\mu\mu}\ran=(\beta(\alpha_s)/16\pi\alpha^2_s) \lan
G^2\ran$ are correspondingly the quark and gluon contributions to
the trace of the energy-momentum tensor.
Note that when deriving this result use was made of the low energy
GMOR relation, and therefore the thermodynamic relation
(\ref{40},\ref{41}) is valid in the light quark theory. Then at
low temperature $T$ when the   excitations of massive hadrons and
interactions of pions can be neglected, equation (\ref{41})
becomes a rigorous QCD theorem. This can be easily verified via
direct calculation using (\ref{28}), (\ref{31}) and the GMOR
relation.

6. The present paper is devoted to the thermodynamic properties of
QCD nonperturbative vacuum with two flavors at low temperature
outside of the scope of perturbation theory. It was shown that at
temperatures $T\la M_\pi$ the relativistic massive pions can be
treated within the dilute gas approximation. Analytic temperature
dependence of the quark condensate perfectly agrees, in the low
temperature region, $T\la M_\pi$, with the
numerical  calculations of $\lan \bar q q\ran(T)$ obtained at the
three-loop level of ChPT with non-zero quark mass \cite{7}. The
gluon condensate slightly varies with the increase of the
temperature, i.e. the situation is similar to the chiral
perturbation theory. It was shown that the temperature derivatives
of the anomalous and normal (quark massive term) contributions to
the trace of the energy-momentum tensor in  QCD with light quarks
are equal to each other in the low temperature region.

As it was mentioned above the pion plays an exceptional role in
thermodynamics of QCD due to the fact that its mass is numerically
close to the phase transition temperature while the masses of
heavier hadrons are several times larger than $T_c$. This was the
reason we did not consider the role of massive states  in the low
temperature phase. This question was discussed in detail in
Ref.\cite{7}. It was shown there that at low temperatures, the
contribution to $\lan \bar q q\ran$ generated by the massive states
is very small, less than 5\% if $T$ below 100 MeV. At $T=150$ MeV,
this contribution is of the order of 10\%. The influence of
thermal excitations of massive hadrons on the properties of the
gluon and quark condensates in the framework of the
conformal-nonlinear $\sigma$- model was also studied in detail in
\cite{21}.

I am grateful to V.A.Rubakov and Yu.A.Simonov for useful
discussions and comments. The financial support of RFFI grant
00-02-17836 and INTAS grant CALL 2000 N 110 is gratefully acknowledged.

  \end{document}